\documentclass[ASNA,twocolumn]{USG}

\usepackage{anyfontsize}
\usepackage{graphicx}
\usepackage{amsmath}
\usepackage{amssymb}
\usepackage{amsfonts}
\usepackage{mathtools}
\usepackage{booktabs}
\usepackage{multirow}
\usepackage{array}
\usepackage{makecell}
\usepackage{adjustbox}
\usepackage{xcolor}
\usepackage{colortbl}
\usepackage{enumitem}
\usepackage{float}
\usepackage{algorithm}
\usepackage{algpseudocode}
\usepackage{steinmetz}

\graphicspath{{./}{./images/}}

\makeatletter
\def\NAT@parse@date#1#2#3#4#5#6@@{\def\NAT@year{}\def\NAT@exlab{}}
\makeatother

\makeatletter
\providecommand\algorithmstar@unused{}
\makeatother

\articletype{ORIGINAL ARTICLE}%
\received{}
\revised{}
\accepted{}
\journal{Applied Computational Intelligence and Soft Computing}
\volume{0}
\copyyear{2026}
\startpage{1}
\articledoi{10.1002/0000}

\begin{document}

\title{PromptForSegCXR: Prompt-Driven Multi-Organ and Multi-Disease
Segmentation in Chest X-rays using a Multi-stage Fusion Mechanism}

\author[1]{Abduz Zami}
\author[2]{Shadman Sobhan}
\author[3]{Rounaq Hossain}
\author[3]{Md. Sawran Sorker}
\author[1]{Mohiuddin Ahmed}
\author[4]{Md. Redwan Hossain}
\author[5]{Md Palash Uddin}

\authormark{ZAMI \textsc{et al.}}
\titlemark{PromptForSegCXR: Prompt-Driven Multi-Organ and Multi-Disease
Segmentation in Chest X-rays}

\address[1]{\orgdiv{Department of Computer Science \& Engineering, }%
\orgname{Rajshahi University of Engineering \& Technology, }%
\orgaddress{\city{Rajshahi }\postcode{6204, }\country{Bangladesh}}}

\address[2]{\orgdiv{Department of Electrical and Electronic Engineering, }%
\orgname{Bangladesh University of Engineering and Technology, }%
\orgaddress{\city{Dhaka }\postcode{1000, }\country{Bangladesh}}}

\address[3]{\orgdiv{Sher-E-Bangla Nagar, }%
\orgname{Shaheed Suhrawardy Medical College, }%
\orgaddress{\city{Dhaka }\postcode{1207, }\country{Bangladesh}}}

\address[4]{\orgdiv{Department of Biomedical Engineering, }%
\orgname{Bangladesh University of Engineering and Technology, }%
\orgaddress{\city{Dhaka }\postcode{1000, }\country{Bangladesh}}}

\address[5]{\orgdiv{Department of Computer Science and Engineering, }%
\orgname{Hajee Mohammad Danesh Science and Technology University, }%
\orgaddress{\city{Dinajpur }\postcode{5200, }\country{Bangladesh}}}

\corres{Md Palash Uddin (\email{palash\_cse@hstu.ac.bd})}


\abstract[ABSTRACT]{Image segmentation is central to automated medical image analysis, enabling precise identification of anatomical structures and pathological regions. Conventional segmentation models typically target a single organ or disease, limiting their adaptability across clinical scenarios. While multi-organ and multi-disease segmentation has been explored, building such datasets requires extensive manual annotation by medical experts. Prompt-driven segmentation offers a flexible, user-guided alternative that speeds up annotation, yet no prior work has addressed prompt-based interactive segmentation across multiple organs and diseases in chest X-rays. This study makes two main contributions. First, we introduce a novel dataset of expert-designed doodle prompts spanning 23 classes (six organs and seventeen diseases), curated from multiple public chest X-ray datasets for prompt-driven segmentation. Second, we propose PromptForSegCXR, a lightweight dual-input segmentation framework that combines the chest X-ray with user-provided doodle prompts to accurately segment diverse anatomical and pathological regions. The model uses a multi-stage feature fusion strategy to integrate spatial and semantic representations, along with a depthwise-pointwise-residual convolution block with squeeze-and-excitation attention for efficient hierarchical feature extraction and adaptive recalibration. Experimental results show the model achieves a Dice score of 81.62 percent on the full dataset, outperforming SAM-based prompt segmentation models by up to 10 percent and conventional segmentation architectures by up to 23 percent, while remaining lightweight. These results demonstrate the effectiveness of the proposed approach for accurate, flexible, prompt-driven chest X-ray segmentation.}

\abbr{AUC, area under the curve; BN, batch normalization; CNN, convolutional
neural network; CXR, chest X-ray; DPRconvSE, depthwise--pointwise--residual
convolution with squeeze-and-excitation; DW, depthwise convolution; IoU,
intersection over union; PW, pointwise convolution; RC, residual connection;
ROS, random oversampling; SAM, Segment Anything Model; SE,
squeeze-and-excitation; SOTA, state of the art.}

\maketitle
\section{Introduction}
Image segmentation enables localization of pathological regions, enhancing precise and accurate clinical decision-making \cite{aggarwal2011role}. Manual segmentation is time-consuming , inconsistent, and difficult when fine details or small abnormalities are present \cite{cardenas2019advances}. Deep learning-based segmentation methods address these limitations by providing a way for automated segmentation \cite{haque2020deep,seo2020machine,roth2018deep}. In particular, Convolutional Neural Networks (CNNs) have demonstrated a strong ability to learn visual patterns from medical imaging data \cite{miotto2018deep,nisha2021applications,kim2020deep}. These models help to reduce human error and speed up diagnosis workflows, thereby improving patient outcomes \cite{shaikdeep}.

Medical image segmentation relies heavily on annotated datasets. However, manual annotation is a time-consuming and costly process that requires expert clinical knowledge \cite{li2024scribformer}. To alleviate this burden, recent studies have explored automated and weakly supervised approaches, leveraging annotations such as scribbles, doodles, or bounding boxes \cite{valvano2021learning,luo2022scribble,zhang2022cyclemix}. Large-scale foundation models \cite{wang2023large,liang2024foundations} have revolutionized AI and started a new era due to their zero-shot and few-shot generalization across numerous downstream tasks \cite{awais2025foundation}. Among these, the Segment Anything Model (SAM) based fine-tuned models have gained attention \cite{zhang2024segment}.  This introduces prompt-driven picture segmentation with a bounding box, expanding its potential. Due to the basic differences between natural and medical images, medical image segmentation may not be able to employ these base models. Many researchers have contributed several expanded works on SAM for medical image segmentation. Nevertheless, such approaches are often computationally expensive to adapt and primarily rely on coarse prompts such as bounding boxes, single pixel or multiple pixel dots, and others, which may include irrelevant background information and fail to capture fine-grained anatomical details \cite{ma2024segment,cheng2023sam}. In contrast, doodle- or scribble-based prompts provide more precise spatial guidance, making them better suited for irregular and complex structures.

Despite these advancements, most existing works focus on general-purpose medical image segmentation and do not specifically address chest X-ray (CXR) imaging. In particular, there is a lack of prompt-based interactive segmentation methods capable of handling both multi-organ and multi-disease scenarios in CXRs. This represents a significant research gap, given that CXRs are one of the most widely used diagnostic imaging modalities for detecting thoracic conditions such as pneumonia, tuberculosis, pleural effusion, bronchitis and others \cite{mcadams2006recent,subramanian2022review}. Chest X-rays are widely used to classify thoracic diseases such as tuberculosis \cite{wileyYudi} and COVID \cite{wileyFatma}. Segmentation helps isolate the region of interest before classification, thereby enabling more accurate results. Moreover, segmenting disease regions in CXRs is particularly challenging due to their varying size, subtle appearance, and variability across patients.

To address these challenges, we propose a novel prompt-driven interactive segmentation framework in which a user-provided doodle highlights the region of interest, guiding the model to accurately segment the corresponding anatomical or pathological structure. This work reduces the hard work of manual annotation of chest X-rays to a great extent. Considering these challenges, the key contributions of this work are as follows:

\begin{itemize}
\item We present the first doodle-prompt-driven framework for multi-organ and multi-disease segmentation in CXRs.
\item We construct a dataset with expert-designed doodle prompts covering 23 organ and disease classes across multiple datasets and propose a dual-input segmentation model that jointly processes the original CXR and its corresponding doodle prompt.
\item We develop a lightweight segmentation model ($\sim$8M parameters) that outperforms state-of-the-art methods by integrating depthwise separable convolutions, residual connections, Squeeze-and-Excitation (SE) attention, and multi-stage feature fusion for effective hierarchical representation learning.

\end{itemize}

The remainder of this paper is organized as follows. Section \ref{sec:Lit_Rev} reviews related work and identifies research gaps. Section \ref{sec: methodology} describes the dataset and model architecture. Section \ref{sec: exp_setup} outlines experimental settings and evaluation protocols. Section \ref{sec:result} presents results and comparative analyses. Section \ref{sec:conclusions} concludes the study and suggests future research directions.

\section{Literature Review}
\label{sec:Lit_Rev}
Image segmentation is a basic computer vision technique that helps to separate areas of interest from the background. It works with classification and object detection. Region growing was one of the first ways to segment an image. It starts with a seed pixel and adds neighboring pixels with similar properties (like color or intensity) over and over again until it forms a uniform region. Many researchers have explored various classification and image retrieval strategies for CXR analysis \cite{baltruschat2019comparison, haq2021deep}. Image segmentation could help reduce unnecessary contextual information and improve the focus on relevant pathological regions. Zucker \cite{zucker1976region} reviewed region-growing systems and their significance in image segmentation. Other classical approaches included thresholding \cite{kohler1981segmentation}, Canny edge detection \cite{canny1986computational}, and clustering-based segmentation \cite{coleman1979image}. Although effective in early applications, these traditional methods were largely replaced by deep learning–based approaches due to their limited adaptability and performance on complex medical images.

The invention of CNNs by LeCun et al. \cite{lecun1998gradient} revolutionized computer vision and paved the way for fully convolutional architectures. Long et al. \cite{long2015fully} introduced the Fully Convolutional Network (FCN), establishing an end-to-end framework for semantic segmentation. This laid the foundation for the U-Net architecture proposed by Ronneberger et al. \cite{ronneberger2015u}, which became the cornerstone of medical image segmentation.

Later improvements led to models like U-Net++ \cite{unet++}, V-Net \cite{milletari2016v}, Attention U-Net \cite{oktay2018attention}, Swin-UNet \cite{swin}, DeepLab \cite{chen2017deeplab}, Mask R-CNN \cite{he2017mask}, and SegNet \cite{badrinarayanan2017segnet}, which were usually made for medical image segmentation for single-class or multiclass and are not capable of utilizing doodles or scribble prompts. Recent work has combined attention mechanisms and generative paradigms to make feature representation and generalization better on small amounts of medical data. Cheng et al. \cite{cheng2024gu} came up with GU-Net, which combines counterfactual attention with a GAN discriminator to make features more varied and less likely to overfit. Tang et al. \cite{tang2024fssn} used Global–Local Aggregation and Feature Filter modules to combine features from the spatial and frequency domains, which made the boundary segmentation sharper. Cheng et al. \cite{cheng2024amnnet} also created AMNNet, which uses Residual U-CBAM modules to learn from multiple scales and reduce noise. These studies all show that hybrid attention, multi-scale, and frequency-aware methods are becoming more common in medical image segmentation.

For multi-organ segmentation, Liu et al. \cite{liu2019automatic} utilized a fully convolutional DenseNet with pixel-weighted loss to automatically delineate ribs and clavicles. Wang et al. \cite{wang2020mdu} put forward the Multitask Dense Connection U-Net (MDU-Net) for simultaneous segmentation of the clavicles, ribs, and lungs, incorporating a mask encoding mechanism to enhance the capture of background features. Jie et al. \cite{wang2019instance} utilized Mask R-CNN for instance segmentation of lung fields, heart, and clavicles, whereas Kholiavchenko et al. \cite{kholiavchenko2020contour} enhanced segmentation accuracy by incorporating organ contour information. Wang et al. \cite{wang2024multi} introduced a multi-objective segmentation technique utilizing collaborative learning from various partially annotated datasets. These methods show how far multi-organ segmentation has come, but they don't allow for interaction, which is an important part of flexible, clinician-led applications. Recently, there has also been a lot of interest in organ segmentation methods that are specific to CXRs. Likewise, Zami et al. \cite{zami2025tig} presented a thresholded input-guided modified UNet++ architecture for precise lung region delineation in CXRs.

Segmentation is also very important for finding the exact location of a disease. Havaei et al. \cite{havaei2017brain} showed how to use it to look at brain tumors, and Campello et al. \cite{campello2021multi} shared results from the M\&Ms Challenge on multi-disease cardiac segmentation. Training these models requires a quality dataset, and creating a dataset for medical image segmentation remains a time-consuming task and requires medical professionals. Prompt-driven segmentation has recently become a new way of segmenting images. It reduces the manual labour of medical experts and opens a way for interactive medical image segmentation. Models such as CLIP \cite{radford2021learning}, ALIGN \cite{jia2021scaling}, Flamingo \cite{alayrac2022flamingo}, and BLIP \cite{li2022blip} amalgamate visual and textual modalities. Meta AI's SAM \cite{kirillov2023segment} revolutionized segmentation by facilitating interactive mask generation through prompts like points, bounding boxes, or textual descriptions. However, SAM's performance in medical imaging is still not great. This has led to domain-specific adaptations like SAM-Med2D \cite{cheng2023sam}, which was fine-tuned on 4.6 million images, and the Medical SAM Adapter (Med-SAM) \cite{wu2023medical}, which adds medical priors through lightweight adaptation. MedSAM, a basic model for universal medical image segmentation, was introduced by Jun et al. \cite{ma2024segment}. Luo et al. \cite{luo2023segclip} created SegCLIP, which uses text prompts for semantic segmentation. Alsaedi et al. \cite{alsaedi2025prompt} developed a Prompt-driven multimodal segmentation framework that employs dynamic fusion to improve the robustness and adaptability of medical imaging for accurate cancer diagnosis through text prompts. Scribble and doodle-based activities have recently garnered significant attention.  Wu et al. \cite{wu2024one} put forward One-Prompt Segmentation, which combines one-shot and interactive paradigms. Wong et al. \cite{wong2024scribbleprompt} created ScribblePrompt, a UNet-based model that can use doodles or scribbles to separate any type of biomedical image.

Even with these improvements, there are still large gaps. Current models seldom tackle interactive multi-organ and multi-disease segmentation in CXRs. Current studies are either focused on specific organs or restricted to a limited number of disease categories, and there is no specialized dataset available for prompt-driven segmentation encompassing all organs and diseases in chest radiography. The proposed work addresses these deficiencies by introducing PromptForSegCXR and a newly curated doodle-prompt dataset.

\section{Methodology} 
\label{sec: methodology}

This section describes the proposed PromptForSegCXR framework, which performs Prompt-driven segmentation of multiple organs and diseases in CXRs using a dual-input architecture. The model receives both the raw CXR and its corresponding doodle prompt to produce an accurate segmentation mask. The key methodological components include dataset preparation, prompt generation, pre-processing, and model design.

\begin{figure*}[!t]
    \centering
    \includegraphics[width=0.92\linewidth]{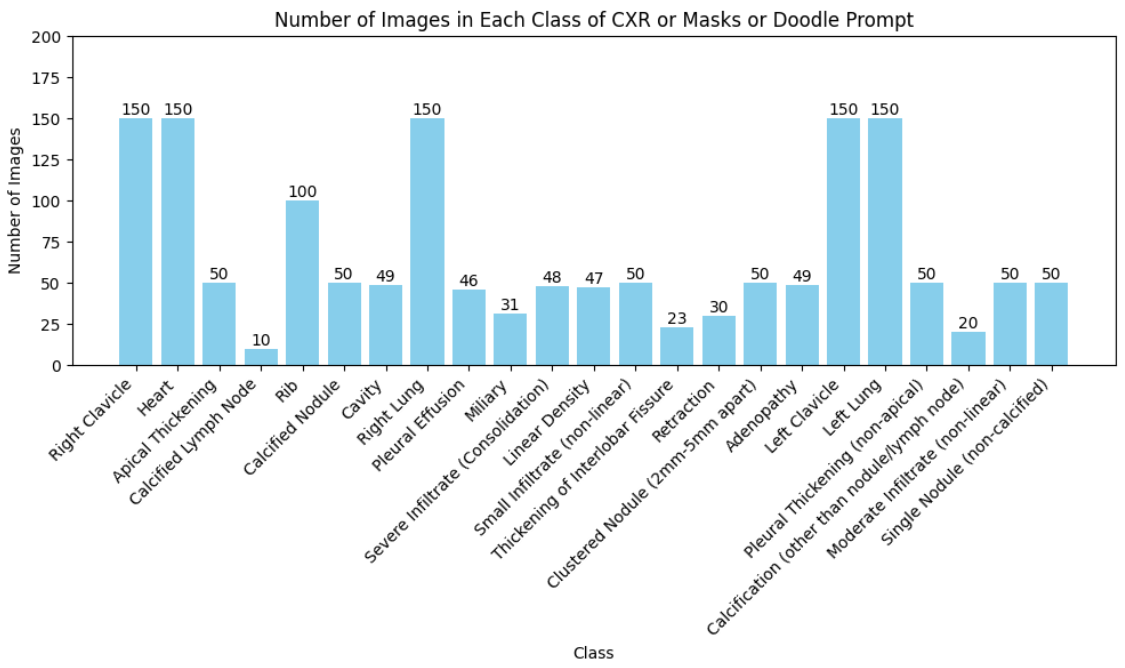}  
    \caption{Number of images in each class of CXR or masks, or doodle prompt}
    \label{fig:ImagevsClass}
\end{figure*}

\subsection{Dataset Description}

CXRs were compiled from three publicly available datasets.

\begin{itemize}
    \item The JSRT dataset \cite{JSRT} provided by the Japanese Society of Radiological Technology consists of 247 postero-anterior chest radiographs with a resolution of \(2048 \times 2048\) pixels and 12-bit grayscale depth. Among these, 154 images contain lung nodules while 93 are nodule-free. Although the original JSRT dataset does not include segmentation masks, organ-wise ground truth masks for the right clavicle, left clavicle, heart, right lung, and left lung were sourced from Ginneken et al. \cite{bram_van_ginneken_2022_7056076}. A total of 150 images per class were selected for this experiment involving these anatomical structures.
    
    \item The VinDr-RibCXR dataset \cite{nguyen2021vindr} is a specialized benchmark designed for automatic segmentation and labeling of individual ribs in CXRs. It comprises 245 chest radiographs with expert-annotated ground truth masks for 20 individual ribs (10 left and 10 right). The annotations were performed by experienced radiologists. For the current study, 100 CXR images along with their corresponding rib segmentation masks were utilized to support bone-related analysis tasks.
    
    \item The Shenzhen Hospital dataset \cite{Shenzhen} collected in collaboration with the U.S. National Library of Medicine contains 662 frontal CXRs, including 326 normal cases and 336 cases exhibiting various manifestations of tuberculosis. Disease-specific segmentation masks for tuberculosis-related abnormalities (such as infiltrates, cavities, and nodules) were employed in this work, with 50 samples selected per disease class to facilitate targeted abnormality segmentation experiments. Few classes had fewer than 50 samples; all the available samples were taken in that case.
\end{itemize}

In total, the dataset included 23 classes—6 organs (Right Clavicle, Left Clavicle, Right Lung, Left Lung, Heart, and Rib) and 17 disease-related regions, including apical thinning, pleural effusion, miliary, retraction, and others. Each sample comprised a CXR, a doodle prompt, and a corresponding segmentation mask. The number of images belonging to each class, is shown in Figure \ref{fig:ImagevsClass}.

\subsection{Prompt Preparation}

Medical experts created doodle prompts using Microsoft Image Viewer by marking regions of interest corresponding to specific organs or disease-affected areas. Each doodle was generated using a single uniform color, without any class-specific color coding. The annotated pixels were then isolated, and all background pixels were set to zero to ensure clear separation of the prompt from the image. To maintain consistency and prevent unintended encoding of class information, all prompts were converted to grayscale and normalized to the range \([0,1]\). This ensures that class identity cannot be inferred from pixel intensity values. Additionally, the doodles consist of sparse, free-form strokes with non-uniform and randomly varying line widths, designed to mimic realistic user input rather than encode semantic labels. The resulting dataset, including mapping information and documentation, is publicly available at
\href{https://data.mendeley.com/datasets/mk36vt2nzj/1}{https://data.mendeley.com/datasets/mk36vt2nzj/1} \cite{zami2024prompt2segcxr}.
 

\subsection{Pre-Processing}

To ensure uniformity and improve segmentation accuracy, several preprocessing operations were applied:

\begin{itemize}
    \item \textbf{Cropping:} Bounding boxes were computed based solely on the non-zero regions of the doodle prompt (i.e., user-provided input) with a fixed padding of 100 pixels to preserve sufficient contextual information. Importantly, ground-truth segmentation masks were \textbf{not} used in this process to avoid any potential information leakage. Cropping was skipped when the region of interest covered more than half of the image area.
    \item \textbf{Resizing and Grayscale Conversion:} All images and prompts were resized to \(256 \times 256\) pixels and converted to grayscale to ensure consistency across datasets.
    \item \textbf{Histogram Equalization:} Histogram equalization was applied to enhance the distribution of pixel intensities and improve contrast, thereby increasing the visibility of anatomical structures and subtle disease patterns.
    \item  \textbf{Normalization:} All images were normalized to the range \([0, 1]\) by dividing pixel intensities by 255. This step ensured consistent intensity scaling across images from different datasets and acquisition devices, stabilized the training process, accelerated convergence, and mitigated gradient-related issues during optimization.
    
\end{itemize}

\subsection{Model Architecture}
\begin{figure}[!t]
     \hspace{0pt}
    \centering
    \includegraphics[width=\linewidth]{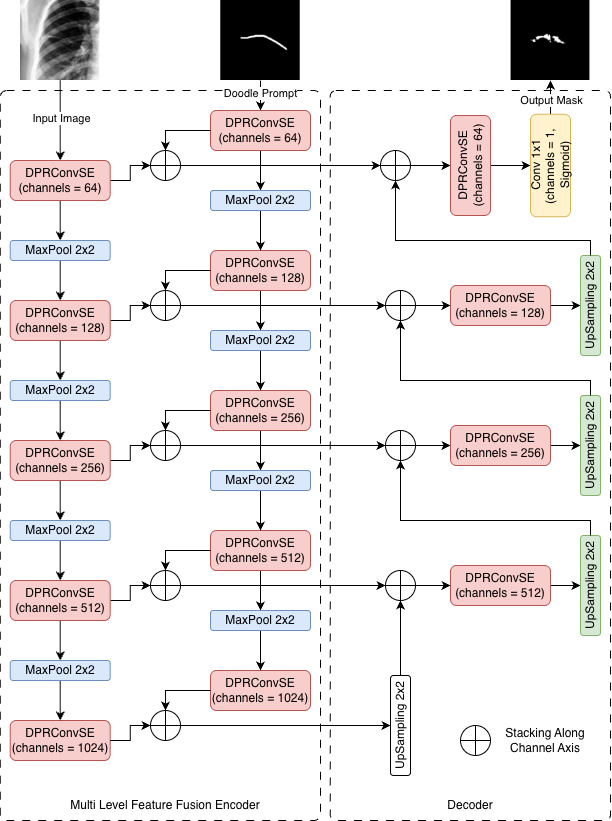}  
\caption{Overview of the proposed PromptForSegCXR architecture. The model adopts a dual-input encoder–decoder framework that processes both the CXR and the corresponding doodle prompt. Multi-stage feature fusion is performed at different encoder levels to effectively integrate spatial and semantic information from both inputs, enabling accurate segmentation of organs and disease regions.}    \label{fig:PromptForSegCXR}
\end{figure}

The suggested PromptForSegCXR has a dual-input and dual-branch encoder-decoder structure that works with both the original CXR and the doodle prompt. The model has a custom DPRconvSE block that uses SE attention and multi-level feature fusion to find both semantic and spatial correlations. The architecture is described below and shown in Figure~\ref{fig:PromptForSegCXR}.

\subsubsection{DPRconvSE Block}
\begin{figure*}[!t]
     \hspace{0pt}
    \centering
    \includegraphics[width=0.85\linewidth]{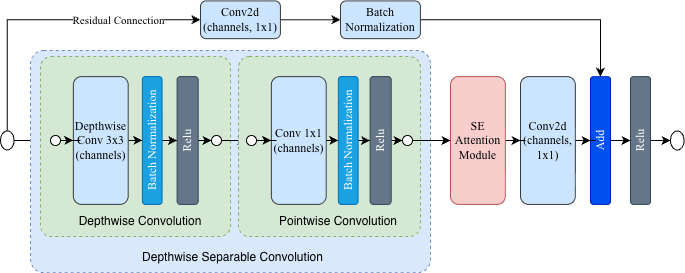}  
\caption{Architecture of the proposed DPRconvSE block. The block combines depthwise and pointwise convolutions for efficient feature extraction, incorporates a residual connection to facilitate gradient flow, and applies channel-wise attention through the SE module to enhance informative features.}    \label{fig:dprconvse}
\end{figure*}

The DPRconvSE block, shown in Figure \ref{fig:dprconvse}, has a 3×3 depthwise convolution, batch normalization, and ReLU activation, followed by a 1×1 pointwise convolution and a SE module. A 1×1 convolution passes through a residual skip connection to make training more stable. The block gets high-level representations quickly and easily, with very little work. The transformation within the DPRconvSE block for an input feature map $\mathbf{X} \in \mathbb{R}^{H \times W \times C}$ is defined through a sequence of operations. First, the depthwise convolution ($F_{dw}$) and pointwise convolution ($F_{pw}$) stages are applied:
\begin{equation}
    \mathbf{X'_{dw}} = \sigma(\text{BN}(\text{Conv}_{dw}(\mathbf{X})))
\end{equation}
\begin{equation}
    \mathbf{X'_{pw}} = \sigma(\text{BN}(\text{Conv}_{pw}(\mathbf{X'_{dw}})))
\end{equation}
where $\sigma$ denotes the ReLU activation and \text{BN} represents Batch Normalization. The pointwise output is then recalibrated by the SE module ($\text{F}_{se}$) and integrated with the residual connection:
\begin{equation}
    \text{F}_{DPRconvSE}(\mathbf{X}) = \sigma(\text{F}_{se}(\mathbf{X'_{pw}}) + \text{Conv}_{1 \times 1}(\mathbf{X}))
\end{equation}
Here, the $1 \times 1$ convolution in the residual path ensures channel consistency between the input and the processed features.

\subsubsection{SE Attention Module}
\begin{figure}[!t]
     \hspace{0pt}
    \centering
    \includegraphics[width=\linewidth]{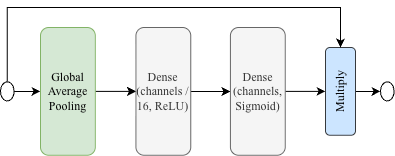}  
\caption{Structure of the SE attention module. The module performs global average pooling followed by two fully connected layers to learn channel-wise importance weights, which are then used to recalibrate feature maps and improve the model’s focus on relevant regions.}    
\label{fig:se}
\end{figure}

Following Hu et al. \cite{hu2018squeeze}, the SE attention module, illustrated in Figure \ref{fig:se}, applies global average pooling followed by two fully connected layers (with ReLU and sigmoid activations) to generate channel-wise weights, which are then multiplied with the input feature map for adaptive recalibration. The SE module performs adaptive channel-wise feature recalibration. Given the input $\mathbf{X'_{pw}}$, the squeeze operation $\boldsymbol{z} \in \mathbb{R}^C$ is calculated via Global Average Pooling:
\begin{equation}
    \boldsymbol{z} = \frac{1}{H W} \sum_{i=1}^{H} \sum_{j=1}^{W} \mathbf{X}'_{pw}(i, j, c)
\end{equation}
The excitation operation then computes the channel weights $\boldsymbol{s}$:
\begin{equation}
    \boldsymbol{s} = \sigma \left( \mathbf{W}_2 \, \delta \left( \mathbf{W}_1 \boldsymbol{z} \right) \right)
\end{equation}
where $W_1 \in \mathbb{R}^{\frac{C}{r} \times C}$ and $W_2 \in \mathbb{R}^{C \times \frac{C}{r}}$ are weights of the two dense layers with a reduction ratio $r=16$. The final output of the SE block is:
\begin{equation}
    \text{F}_{se}(\mathbf{X'_{pw}}) = \boldsymbol{s} \cdot \mathbf{X'_{pw, c}}
\end{equation}

\subsubsection{Encoder}

The encoder gets two single-channel inputs: the original CXR and the doodle prompt that goes with it. Both inputs go through the same feature extraction pipelines. In the first stage, each input goes through a DPRconvSE block, outputting 64 channels. Then, a fusion block combines the feature maps that come out of the block. The fused features are then passed through four more stages, each of which has a 2×2 max-pooling layer and a DPRconvSE block. The number of output channels in the DPRconvSE block doubles at each stage, until it reaches 1024 in the last stage. Dedicated fusion blocks are used to combine the outputs from the CXR and prompt branches after each DPRconvSE operation. This makes sure that both branches learn from each other. At each encoder stage $i \in \{1, \dots, 5\}$, the feature maps from the CXR branch $\text{F}_{DPRconvSE}(\textbf{X}_{cxr},_{i})$ and the prompt branch $\text{F}_{DPRconvSE}(\textbf{X}_{prompt},_{i})$ are merged using a fusion operator $\Phi$:
\begin{equation}
    \mathbf{Fusioned}_i = \Phi({\text{F}_{DPRconvSE}(\textbf{X}_{cxr},_{i})}, \text{F}_{DPRconvSE}(\textbf{X}_{prompt},_{i}))
\end{equation}
\begin{equation}
    \mathbf{Fusioned}_i = [{\text{F}_{DPRconvSE}(\textbf{X}_{cxr},_{i})}\oplus \text{F}_{DPRconvSE}(\textbf{X}_{prompt},_{i})]
\end{equation}
where $\oplus$ represents stacking one over another through the channel axis. This fused map $\mathbf{Fusioned}_i$ is then propagated to the subsequent max-pooling layer and the corresponding decoder level $i$ via skip connections. The encoder finally makes five fused feature maps, one from each stage. These maps are then sent to the decoder to be put back together. The encoder uses DPRconvSE blocks to extract features in a hierarchical way, 2×2 max pooling to downsample spatially, and multi-level fusion to combine representations from both the CXR and doodle prompt branches. This design makes it possible to encode both semantic and spatial information at the same time before decoding.

\subsubsection{Decoder}

The segmentation mask is created by the decoder, which reassembles the five fused encoder outputs. Starting from the deepest feature map (stage 5), a 2×2 kernel was employed to upsample the features, which were subsequently combined with the features from the stage 4 fusion block:
\begin{equation}
\mathbf{DecFusioned}_4 =
\Phi(
\text{US}(\mathbf{Fusioned}_5),
\ \mathbf{Fusioned}_4)
\end{equation}
where $\mathbf{DecFusioned}_i$ is the fusion output at each decoder stage, and US is the upsampling function.
Subsequently, this combined output was refined by a DPRconvSE block that converted the feature map to 512 channels and upsampled. The decoder consists of three more stages. At each stage, the upsampled features are combined with those from the preceding fusion block and transmitted through a DPRconvSE block. The spatial resolution is progressively increased by 2×2 upsampling, and the number of channels is reduced by half as approaching the output layer. This process can be represented as:
\begin{equation}
\begin{split}
\mathbf{DecFusioned}_i = \Phi\big(
&\text{US}(\mathbf{F}_{\text{DPRconvSE}}(\mathbf{DecFusioned}_{i+1})),\\
&\ \mathbf{Fusioned}_i\big), \quad i = 3,2,1
\end{split}
\end{equation}
In the final stage of decoding, the upsampled features are combined with those from the stage 1 fusion block and sent through a DPRconvSE block with 64 output channels, followed by a 1×1 convolution. Next, the final segmentation mask $\mathbf{M}$ is generated using a sigmoid activation function. 
\begin{equation}
\mathbf{M} =
\sigma \Big(
\text{Conv}_{1 \times 1} (
\mathbf{F}_{\text{DPRconvSE}}(\mathbf{DecFusioned}_1)))
\end{equation}
Consequently, the decoder concatenates and upsamples multi-scale contextual features. It enhances spatial accuracy and feature reconstruction by employing DPRconvSE blocks. The proposed PromptForSegCXR model is generally explained in the algorithm~\ref{alg:PromptForSegCXR}.

\begin{algorithm}[!t]
\small
\caption{PromptForSegCXR Model with Dual Inputs}
\label{alg:PromptForSegCXR}
\begin{algorithmic}[1]
\State \textbf{Input:}
\State \quad $I$: Image Input of shape $(256, 256, 1)$.
\State \quad $D$: Doodle Input of shape $(256, 256, 1)$.
\State \textbf{Output:}
\State \quad $M$: Segmentation mask of shape $(256, 256, 1)$.

\State \textbf{Step 1: Define DPRconvSE Block}
\State \quad \textbf{Input:} Feature map $x$, output channels $f$, kernel size $k=3$, activation $\sigma=\text{ReLU}$.
\State \quad Apply depthwise convolution with $k \times k$ kernel and batch normalization.
\State \quad Apply pointwise convolution with $f$ output channels and batch normalization.
\State \quad Apply SE Block:
\State \quad \quad Compute global average pooling on $x$.
\State \quad \quad Dense layer with $f / 16$ units and ReLU activation.
\State \quad \quad Dense layer with $f$ units and sigmoid activation.
\State \quad \quad Multiply SE weights with $x$.
\State \quad Add a residual connection between input $x$ and SE output.
\State \quad \textbf{Output:} Transformed feature map.

\State \textbf{Step 2: Define Image and Doodle Encoder}
\State \quad \textbf{Input:} Input image $X$ of shape $(256, 256, 1)$.
\State \quad At each scale, apply the
\State \quad \quad DPRconvSE block with output channels: $[64, 128, 256, 512, 1024]$.
\State \quad \quad MaxPooling operation for downsampling.
\State \quad \textbf{Output:} Feature maps from multiple scales: $\text{conv1}, \text{conv2}, \text{conv3}, \text{conv4}, \text{conv5}$.

\State \textbf{Step 3: Fuse Features from Image and Doodle Encoders}
\State Extract feature maps $\text{conv1--conv5}$ from both $I$ and $D$.
\State Fuse feature maps at each scale using element-wise concatenation:
\State \quad $\text{fusion1} = \text{Concat}(\text{image\_conv1}, \text{doodle\_conv1})$.
\State \quad $\text{fusion2} = \text{Concat}(\text{image\_conv2}, \text{doodle\_conv2})$.
\State \quad $\dots$

\State \textbf{Step 4: Decoder Path (Upsampling and Skip Connections)}
\State \quad Start with a fused feature map $\text{fusion5}$:
\State \quad \quad Apply upsampling by $2 \times 2$.
\State \quad \quad Concatenate with $\text{fusion4}$ (skip connection).
\State \quad \quad Apply DPRconvSE with $512$ output channels.
\State \quad Repeat for $\text{fusion3}, \text{fusion2}, \text{fusion1}$:
\State \quad \quad Upsample, concatenate, and apply DPRconvSE at each scale.

\State \textbf{Step 5: Generate Segmentation Output}
\State \quad Apply a $1 \times 1$ convolution with sigmoid activation on the final decoder layer.
\State \quad Output the segmentation mask $M$.

\end{algorithmic}
\end{algorithm}

\section{Experimental Settings}
\label{sec: exp_setup}

The success of a deep learning model depends heavily on the careful design and execution of its experimental setup. This section outlines the procedures used to prepare, train, and evaluate the proposed segmentation model. These include strategies to mitigate class imbalance, ensure robust validation, tune hyperparameters, define evaluation metrics, and leverage computational resources efficiently. Each component of this setup was essential in ensuring the model’s reliability, efficiency, and accuracy.

\subsection{Handling Class Imbalance}

The dataset was constructed using expert-annotated samples from three publicly available sources: JSRT, VinDr-RibCXR, and the Shenzhen Hospital dataset. The number of images per class was determined based on the availability of annotated data. Specifically, up to 150 images were selected for major organ classes (e.g., lungs, heart, and clavicles), 100 images for the rib class, and up to 50 images per class for the 17 disease categories. For disease classes with limited available annotations (e.g., Calcified Lymph Node and Calcification), all available samples were included to maximize data utilization. To address class imbalance in the dataset, a random oversampling strategy was employed. This approach increases the number of samples in minority classes by randomly duplicating existing instances until all classes contain an equal number of samples. Consequently, the model learns features from all classes more uniformly during training, reducing bias toward majority classes and improving overall performance and generalization. Oversampling was applied only to the training–validation set and not to the test set to ensure a fair and unbiased evaluation.

\subsection{Cross-Validation}

A 5-fold cross-validation strategy was adopted to ensure robustness. The dataset was split at the level of the original CXR, i.e., using group-wise splitting, to prevent any potential data leakage. Specifically, all samples derived from the same CXR (including different prompts or masks) were assigned to the same split. A total of 20\% of the dataset was reserved as the test set, maintaining proportional representation of each class. The remaining 80\% was then partitioned into five equal folds for cross-validation.  For each iteration, one fold served as the validation set while the remaining four were used for training. This process was repeated five times so that each fold served as the validation set exactly once. Each fold was trained for 100 epochs with a batch size of 16. The performance metrics from all folds were averaged to obtain the final evaluation results, providing a reliable estimate of model performance across varying data distributions. 
To further improve generalization and reduce overfitting, data augmentation techniques were applied during training, including random rotation (up to \(12^\circ\)) and random zooming (scaling factor of 0.2).





\subsection{Hyperparameter Tuning}

To achieve optimal performance and stable convergence, several hyperparameters were carefully tuned. The initial learning rate was set to 0.001. A dynamic learning rate reduction strategy was used. If the validation Dice coefficient plateaued for five consecutive epochs, the learning rate was reduced to 20\% of its current value, continuing until it reached a minimum of $1 \times 10^{-9}$. An early-stopping mechanism was implemented to prevent overfitting. Training terminated if the validation Dice coefficient did not improve for 10 consecutive epochs. Additionally, model checkpointing was employed to save the weights corresponding to the highest validation Dice score, ensuring that the best configuration was retained for final evaluation.

\subsection{Evaluation Metrics}

The model’s performance was assessed using four key metrics:

\begin{itemize}
\item Dice Coefficient: Measures overlap between predicted and ground-truth masks \cite{dice}.

\begin{equation} Dice = \frac{2 \cdot |X \cap Y|}{|X| + |Y|} \end{equation}

where $X$ and $Y$ denote the ground-truth and predicted masks, respectively.

\item Jaccard Index (IoU): Quantifies intersection over union between predicted and ground-truth masks \cite{jaccard}.

\begin{equation} Jaccard = \frac{|X \cap Y|}{|X \cup Y|} \end{equation}

\item Area Under the Curve (AUC): Evaluates the model’s discriminative ability by measuring the area under the ROC curve \cite{auc}.

\item Binary Accuracy: Represents the percentage of correctly classified pixels \cite{accuracy}.

\end{itemize}

\subsection{Loss Function} The Dice coefficient loss was employed as the optimization objective, defined as:

\begin{equation} DiceLoss = 1 - DiceCoefficient \end{equation}

Minimizing this loss encourages the model to maximize overlap between predicted and true masks, thus improving segmentation accuracy.

\begin{table*}[!t]
\centering
\caption{Results of Training and Evaluating on Full Dataset}
\begin{tabular*}{\textwidth}{@{\extracolsep\fill}lcccc@{\extracolsep\fill}}
\toprule
Class & Dice & Jaccard & AUC & Accuracy \\ \midrule
Right Clavicle                               & 86.32         & 76.12            & 91.76        & 98.36             \\
Heart                                        & 92.87         & 86.73            & 94.58        & 95.99             \\
Apical Thickening                            & 64.96         & 48.54            & 83.77        & 98.60             \\
Calcified Lymph Node                         & 38.69         & 24.00            & 72.47        & 98.76             \\
Rib                                          & 81.18         & 66.96            & 87.09        & 90.10             \\
Calcified Nodule                             & 76.36         & 65.89            & 85.65        & 99.47             \\
Cavity                                       & 66.32         & 49.68            & 86.29        & 98.61             \\
Right Lung                                   & 95.80         & 91.96            & 96.38        & 98.49             \\
Pleural Effusion                             & 60.55         & 43.83            & 77.81        & 93.81             \\
Miliary                                      & 82.50         & 70.75            & 91.12        & 95.86             \\
Severe Infiltrate (Consolidation)            & 89.61         & 81.31            & 95.17        & 97.24             \\
Linear Density                               & 19.83         & 11.28            & 57.51        & 97.64             \\
Small Infiltrate (non-linear)                & 58.21         & 41.48            & 73.61        & 94.10             \\
Thickening of Interlobar Fissure             & 34.75         & 22.96            & 64.89        & 97.58             \\
Retraction                                   & 44.41         & 31.04            & 68.14        & 93.36             \\
Clustered Nodule (2mm-5mm apart)             & 84.61         & 73.71            & 96.07        & 99.39             \\
Adenopathy                                   & 79.09         & 65.96            & 89.32        & 96.78             \\
Left Clavicle                                & 78.60         & 65.06            & 84.46        & 97.89             \\
Left Lung                                    & 94.09         & 88.95            & 95.01        & 98.32             \\
Pleural Thickening (non-apical)              & 64.66         & 47.87            & 86.99        & 98.64             \\
Calcification (other than nodule/lymph node) & 50.20         & 34.41            & 76.97        & 97.92             \\
Moderate Infiltrate (non-linear)             & 68.40         & 53.60            & 80.52        & 94.94             \\
Single Nodule (non-calcified)                & 39.90         & 27.38            & 71.18        & 98.39             \\
All                                          & 81.62         & 70.38            & 88.92        & 97.03             \\ \bottomrule

\end{tabular*}
\label{tab:full_dataset_train}
\end{table*}

\begin{figure*}[!tbp]
    \centering
    \includegraphics[scale=0.12]{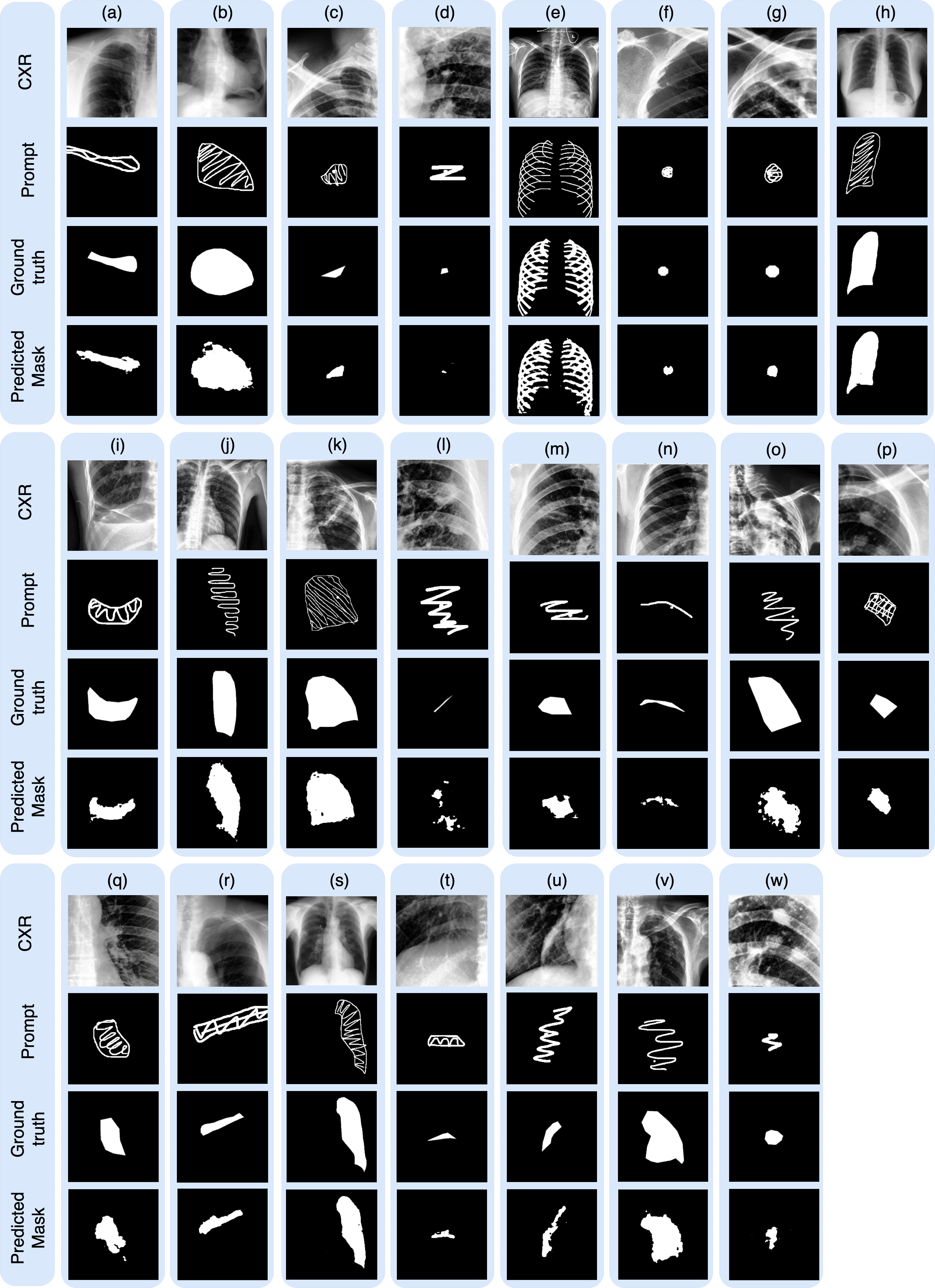}
    \caption{Prediction Samples of Our Model across all the classes (a) Right Clavicle (b) Heart (c) Apical Thickening (d) Calcified lymph node (e) Rib (f) Calcified Nodule (g) Cavity (h) Right Lung (i) Pleural Effusion (j) Miliary (k) Severe Infiltrate (Consolidation) (l) Linear Density (m) Small Infiltrate (non-linear) (n) Thickening of interlobar fissure (o) Retraction (p) Clustered Nodule (2mm-5mm apart) (q) Adenopathy (r) Left Clavicle (s) Left Lung (t) Pleural Thickening (non-apical) (u) Calcification 
(other than the Nodule lymph node) (v) Moderate Infiltrate (non-linear) (w) Single Nodule (non-calcified) }
    \label{fig:Model_performance}
\end{figure*}

\begin{table*}[!tbp]
\centering
\caption{Results of Training and Evaluating on only Organs}
\begin{tabular*}{0.9\textwidth}{@{\extracolsep\fill}lcccc@{\extracolsep\fill}}
\toprule
Class & Dice & Jaccard & AUC & Accuracy \\ \midrule
Right Clavicle & 89.73         & 81.58            & 94.28        & 98.70             \\
Heart          & 94.22         & 89.11            & 96.59        & 98.62             \\
Right Lung     & 97.24         & 94.64            & 98.76        & 98.97             \\
Rib            & 78.92         & 65.21            & 87.15        & 90.35             \\
Left Clavicle  & 91.02         & 83.62            & 95.10        & 98.82             \\
Left Lung      & 97.13         & 94.43            & 98.69        & 99.10             \\
All            & 92.47         & 86.45            & 96.00        & 97.57             \\ \bottomrule
\end{tabular*}
\label{tab:organs}
\end{table*}

\begin{table*}[!tbp]
\centering
\caption{Results of Training and Evaluating on Only Diseases}
\begin{tabular*}{0.8\textwidth}{@{\extracolsep\fill}lcccc@{\extracolsep\fill}}
\toprule
Class                    & Dice & Jaccard & AUC & Accuracy \\ \midrule
Apical Thickening                 & 70.80         & 55.40            & 87.59        & 98.34             \\
Calcified Lymph Node              & 30.14         & 17.75            & 59.67        & 95.90             \\
Calcified Nodule                  & 65.37         & 49.25            & 79.88        & 99.31             \\
Cavity                            & 72.02         & 57.26            & 91.10        & 98.56             \\
Pleural Effusion                  & 62.61         & 45.69            & 80.84        & 94.54             \\
Miliary                           & 81.24         & 68.52            & 88.94        & 95.50             \\
Severe Infiltrate (Consolidation) & 87.53         & 78.46            & 92.68        & 97.12             \\
Small Infiltrate (non-linear)     & 64.42         & 47.91            & 78.32        & 94.72             \\
Retraction                        & 52.73         & 35.89            & 73.58        & 91.20             \\
Clustered Nodule (2mm-5mm apart)  & 77.61         & 63.77            & 92.86        & 98.65             \\
Adenopathy                        & 71.68         & 56.45            & 83.41        & 96.23             \\
Pleural Thickening (non-apical)   & 60.13         & 44.11            & 81.73        & 98.37             \\
Moderate Infiltrate (non-linear)  & 66.05         & 49.85            & 81.03        & 93.67             \\
All                               & 73.08         & 57.63            & 85.63        & 96.86             \\ \bottomrule
\end{tabular*}
\label{tab:diseases}
\end{table*}

\begin{table*}[!b]
    \centering
        \caption{Performance metrics for various model configurations, showcasing the impact of progressively integrating Pointwise Convolution (PW), Depthwise Convolution (DW), SE attention module (SE), Residual Connections (RC), and Random Oversampling (ROS) on segmentation accuracy}
    \begin{tabular*}{\textwidth}{@{\extracolsep\fill}lcccc@{\extracolsep\fill}}
        \toprule
        Cases & Dice & Jaccard & AUC & Accuracy \\
        \midrule
        Baseline               & 66.94 & 51.79 & 87.08 & 93.32 \\
        Baseline+DW+PW               & 78.69 & 66.68 & 91.03 & 96.09 \\
        Baseline+DW+PW+SE            & 79.32 & 67.37 & 90.69 & 96.27 \\
        Baseline+DW+PW+CBAM            & 78.93 & 67.17 & 90.22 & 95.97 \\
        Baseline+DW+PW+ECA            & 79.14 & 67.11 & 90.12 & 96.02 \\
        Baseline+DW+PW+SE+RC         & 79.92 & 68.45 & \textbf{92.42} & 96.33 \\
        Baseline+DW+PW+SE+RC+ROS-PromptBranch         & 69.82 & 61.29 & 76.37 & 95.56 \\
        Baseline+DW+PW+SE+RC+ROS (proposed)     & \textbf{81.62} & \textbf{70.38} & 88.92 & \textbf{97.03} \\
        \bottomrule
    \end{tabular*}
    \label{tab:ablution}
\end{table*}

\subsection{Machine Configuration}

All experiments were conducted on Kaggle’s GPU-accelerated platform \cite{kaggle}, using a notebook environment equipped with two NVIDIA Tesla T4 GPUs (each with 16 GB GDDR6 memory and 2,560 CUDA cores). This configuration provided sufficient computational capacity for training the lightweight PromptForSegCXR model efficiently.

\section{Results \& Discussion}
\label{sec:result}
The proposed model’s performance was evaluated and compared against state-of-the-art (SOTA) segmentation approaches using the Dice, Jaccard, AUC, and accuracy metrics. The model’s ability to generalize across datasets and handle diverse input variations was also analyzed. All scores reported here are the means of scores obtained from cross-validation folds.

\subsection{Scores Obtained by the Proposed Model}
Given the dataset’s composition of both organ and disease categories, three distinct training configurations were tested: (i) the full dataset (organs + diseases), (ii) organs only, and (iii) diseases only. Results for each configuration are summarized in Tables \ref{tab:full_dataset_train}, \ref{tab:organs}, \ref{tab:diseases}. Most classes achieved strong segmentation performance, though a few—such as Linear Density, Thickening of Interlobar Fissure, Calcified Lymph Node, and Single Nodule (non-calcified)—recorded lower scores due to small region sizes and complex boundaries.

\begin{table*}[!b]
\centering
\caption{Comparison of PromptForSegCXR Model and a few SAM-based zero-shot models with additional models trained on our dataset (mean ± std across cross-validation folds)}
\label{tab:Comparison}

\resizebox{\textwidth}{!}{
\begin{tabular}{lcccccccc}
\toprule
Models & Prompt & \makecell{Dice \\ (\%)} & \makecell{Jaccard \\ (\%)} & \makecell{AUC \\ (\%)} & \makecell{Accuracy \\ (\%)} & \makecell{Training} & \makecell{Params \\ (M)} & \makecell{P-value \\ (Dice)} \\ \midrule
SAM \cite{cheng2023sam}        & BBox   & 71.25  & 58.16   & 84.09 & 92.64    & Zero-shot & 91 & 1.39e-16 \\
MedSAM \cite{ma2024segment}    & BBox   & 74.82  & 62.98   & 83.74 & 95.38    & Zero-shot & 91 & 6.26e-10 \\
SAM Med 2D \cite{cheng2023sam} & BBox   & 73.66  & 61.37   & 85.95 & 95.28    & Zero-shot & 91 & 1.56e-12 \\
UNet \cite{ronneberger2015u}      & Overlaid Doodle      & 57.64 $\pm$ 0.19  & 41.12 $\pm$ 0.12   & 76.65 $\pm$ 0.21    & 91.96$\pm$ 0.16        & Fully Trained & 31 & 7.08e-63 \\
UNet++ \cite{unet++}    & Overlaid Doodle      & 57.44$\pm$ 0.21  & 40.88$\pm$ 0.14   & 76.29$\pm$ 0.19     & 91.59$\pm$ 0.12        & Fully Trained & 36.2 & 5.92e-51 \\
Attention UNet \cite{oktay2018attention} & Overlaid Doodle & 62.50$\pm$ 0.14 & 46.00$\pm$ 0.11 & 79.50$\pm$ 0.18 & 93.50$\pm$ 0.14 & Fully Trained & 35.0 & 2.98e-59 \\
Swin-Unet \cite{swin} & Overlaid Doodle & 65.80$\pm$ 0.13 & 49.00$\pm$ 0.09 & 82.10$\pm$ 0.22 & 94.10$\pm$ 0.09 & Fully Trained & 41.4 & 4.32e-43 \\
DeepLabV3 (ResNet) \cite{chen2017rethinking} & Overlaid Doodle & 81.16$\pm$ 0.11 & 70.01$\pm$ 0.09 & 88.44$\pm$ 0.19 & 95.33$\pm$ 0.11 & Fully Trained & 9.0 & 0.3435\\
DeepLabV3+ (ResNet) \cite{chen2018encoder} & Overlaid Doodle & 81.23$\pm$ 0.11 & 70.05$\pm$ 0.14 & 88.52$\pm$ 0.23 & 95.79$\pm$ 0.08 & Fully Trained & 9.7 & 0.3893 \\
ScribblePrompt \cite{wong2024scribbleprompt} & Overlaid Doodle & 72.20$\pm$ 0.13 & 57.28$\pm$ 0.12 & 86.13$\pm$ 0.17 & 94.50$\pm$ 0.12 & Fully Trained & 36.2 & 2.54e-20 \\
Alsaedi et. al. \cite{alsaedi2025prompt} & Overlaid Doodle & 58.22$\pm$ 0.18 & 41.20$\pm$ 0.15 & 76.29$\pm$ 0.24 & 91.18$\pm$ 0.13 & Fully Trained & 8.0 & 2.25e-69 \\
Ours       & Doodle & \textbf{81.62$\pm$ 0.12} & \textbf{70.38$\pm$ 0.11} & \textbf{88.92$\pm$ 0.17} & \textbf{97.03$\pm$ 0.09} & Fully Trained & \textbf{7.2} \\
\bottomrule
\end{tabular}}

\end{table*}

PromptForSegCXR functions more efficiently as a result of its innovative construction. Depthwise and pointwise convolutions facilitated the extraction of features without necessitating an excessive amount of processing capacity. The gradient's disappearance was prevented by residual connections, which also preserved critical features. SE attention modules concentrated on specific spatial regions. Additionally, the multi-stage fusion mechanism effectively combined high-level semantic context with low-level texture information, resulting in segmentation that was both stable and accurate in a variety of image conditions.

Even though it possessed certain advantages, the model performed better at organ segmentation than disease segmentation. The small and irregular shapes of numerous disease-affected regions complicate the process of creating illustrations by hand and instructing models. Additionally, a few minor misclassifications were the result of class imbalance and overlapping visual patterns, such as those between pleural effusion and adenopathy. The results were also influenced by the doodle prompts; prompts that were not as plain occasionally resulted in less accurate results. Nevertheless, the model demonstrated the ability to generalize effectively and operate efficiently across a variety of tasks, as illustrated in Figure \ref{fig:Model_performance}.

To further assess model reliability, the percentage error was computed for each class as \(100\% - \text{Dice}\). The results indicate low error rates for well-defined anatomical structures, such as the Right Lung (4.2\%), Left Lung (5.91\%), and Heart (7.13\%). The model also demonstrates strong performance on several disease categories, including Severe Infiltrate/Consolidation (10.39\%) and Clustered Nodules (15.39\%). However, higher error rates are observed in more challenging pathologies, such as Linear Density (80.17\%) and Thickening of Interlobar Fissure (65.25\%), which are characterized by small size, elongated shapes, and subtle visual boundaries. These results highlight the model’s robustness in segmenting well-defined regions while also reflecting the inherent difficulty of accurately delineating fine-grained pathological structures.

\subsection{Ablation Study}

Ablation experiments were conducted to evaluate the individual and cumulative effects of the proposed architectural components and techniques. Starting from a baseline network with multi-stage fusion, we sequentially incorporated pointwise and depthwise convolutions, SE attention, residual connections, and random oversampling for class balance. The outcomes are shown in Table \ref{tab:ablution}. The baseline model achieved a Dice score of 66.94\% and a Jaccard index of 51.79\%. Adding depthwise convolutions improved feature extraction efficiency, raising the Dice score to 78.69\%. Incorporating SE attention further enhanced focus on salient regions, increasing the Dice to 79.32\%. CBAM and ECA mechanisms were also tested but performance dropped. Introducing residual connections facilitated gradient flow, achieving 79.92\%. Finally, applying random oversampling yielded the best performance— 81.62\% Dice, and 70.38\% Jaccard. Model's performance without the prompt extraction part was also tested and it significantly reduced the performance. These results confirm that the integration of advanced architectural modules with balanced data significantly strengthens segmentation performance.

\subsection{Comparison with Other Work}
To our knowledge, this is the first study on Prompt-driven chest radiograph segmentation. Since no prior models were trained on this dataset, we compared PromptForSegCXR with representative variants of SAM, a leading general-purpose Prompt-driven segmentation framework that uses bounding-box prompts. Additionally, several SOTA medical segmentation models—UNet, UNet++, Attention UNet, Swin-UNet, DeepLabV3, and DeepLabV3+—were retrained in our dataset along with specialized Prompt driven segmentation models such as ScribblePrompt and the proposed model by Alsaedi et. al. As these models accept single-image input, the doodle prompt was overlaid on the chest radiograph to form a three-channel RGB input. Standard deviation was not possible to add for SAM based models as they were just fined tuned on our dataset. 

Table \ref{tab:Comparison} compares PromptForSegCXR (trained on the full dataset) with SAM-based medical segmentation models (evaluated in zero-shot mode) and traditional models (trained on the full dataset). The proposed model consistently outperformed both SAM-based and conventional architectures, demonstrating higher accuracy and Dice scores despite its compact 8 M parameter size. Cross-validation was performed to ensure the robustness and generalizability. However, cross-validation could not be conducted on SAM-based models, as they require significant computational resources and take a long time to train. The doodle-based prompts provided richer spatial cues than the bounding boxes, enabling finer boundary delineation. In contrast, traditional SOTA models underperformed because they were not optimized to process prompt overlays. ResNet based DeepLabV3 and DeepLabV3+ gave a good competition with ours. PromptForSegCXR achieved comparable performance to DeepLabV3 and DeepLabV3+ while significantly outperforming all other models (p<0.001). The success of PromptForSegCXR primarily stems from its dual-input multi-stage fusion mechanism, which effectively integrates semantic information from both the CXR and the user-provided prompt.

\section{Conclusion and Future Work}
\label{sec:conclusions}

This study presents an innovative dataset of doodle-based prompts for CXR segmentation, featuring six organs and seventeen diseases. It also introduces a lightweight model, PromptForSegCXR, intended for interactive, prompt-guided segmentation. PromptForSegCXR is different from traditional single-task segmentation models because it allows for flexible, multi-organ, and multi-disease segmentation with high accuracy while still being fast enough for use in clinical settings. The model's experimental results show that it is more accurate and robust than the best methods currently available, setting a new standard for Prompt-driven medical image segmentation. Even though it works well overall, the accuracy of segmentation for some diseases is still a little lower than for organs. This is mostly because diseased areas are small or have irregular shapes, there is an imbalance in the number of classes, and it is hard to make accurate doodles. Finding ways to get around these problems is a big part of future work. Possible extensions include adding CT, MRI, and ultrasound images to the modality coverage to see if the model can generalize; using transfer learning to make the model more accurate for certain diseases or clinical settings; and adding PromptForSegCXR to clinical workflows, which would let radiologists do real-time interactive segmentation to help them. By solving these problems, future research can improve Prompt-driven segmentation even more, making medical imaging diagnostic tools that are more flexible, accurate, and easy to understand.


\bmsubsection*{Acknowledgments}

The authors thank the medical experts who prepared the doodle prompts used in this study, and gratefully acknowledge the providers of the JSRT, VinDr-RibCXR and Shenzhen Hospital chest radiograph collections. A preprint has previously been published in \cite{zami2025prompt2segcxrpromptsegmentorgansdiseases}.

\subsection*{Declaration of interests}
The authors declare that they have no known competing financial interests or personal relationships that could have appeared to influence the work reported in this paper.

\subsection*{Funding sources}
This research did not receive any specific grant from funding agencies in the public, commercial, or not-for-profit sectors.

\subsection*{Declaration of AI} 
During the preparation of this work, the authors used GenAI to improve grammar and spelling. After using this tool/service, the authors reviewed and edited the content as needed and took full responsibility for the content of the publication.

\bmsubsection*{Data Availability Statement}

The doodle-prompt dataset generated in this study, together with the mapping information and documentation required to align it with the source chest X-ray collections, is publicly available at
\url{https://data.mendeley.com/datasets/mk36vt2nzj/1}.

\subsection*{Ethics Approval and Consent to Participate}

 This study did not involve the recruitment of human participants, direct interaction with individuals, or the collection of identifiable human data by the authors. All experiments were conducted exclusively using the publicly available chest X-ray collections \url{https://data.mendeley.com/datasets/mk36vt2nzj/1}. These datasets were previously collected, de-identified, and released for research purposes by their respective providers. Consequently, this study did not require additional institutional ethics committee or institutional review board approval, nor informed consent from participants. The research was conducted in accordance with the terms of use of the respective datasets.

\bibliography{cas-refs}

\end{document}